\documentclass[useAMS,usenatbib]{mn2e}
\usepackage[dvips]{graphicx}
\usepackage{txfonts}
\usepackage{amssymb}
\usepackage{indentfirst}

\begin{document}

\title{Unresolved emission and ionized gas in the bulge of M31}

\author[\'A.Bogd\'an \& M.Gilfanov]{\'A.Bogd\'an$^{1}$\thanks{E-mail:
bogdan@mpa-garching.mpg.de; gilfanov@mpa-garching.mpg.de} 
and
M. Gilfanov$^{1,2}$\footnotemark[1]\\
$^{1}$Max Planck Institut f\"ur Astrophysik, Karl-Schwarzschild-Str.1,
85741 Garching bei M\"unchen, Germany\\
$^{2}$Space Research Institute, Russian Academy of Sciences, Profsoyuznaya
84/32, 117997 Moscow, Russia}

\date{}

\maketitle

\begin{abstract}
We study the origin of unresolved X-ray emission from the bulge 
of M31 based on archival Chandra and XMM-Newton observations. 
We demonstrate that three different components are present: 
(i) Broad-band emission from a large number of faint sources -- mainly 
accreting white dwarfs and active binaries, associated with the old
stellar population, similar to the Galactic Ridge X-ray emission of
the Milky Way. The X-ray to K-band luminosity ratios are compatible
with those for the Milky Way and for M32, in the $ 2 - 10 \
\mathrm{keV} $ band it is $ (3.6 \pm 0.2) \times 10^{27} \ \mathrm{erg 
\ s^{-1} \ L_{\sun}^{-1}} $. 
(ii) Soft emission from ionized gas with
temperature of about $\sim 300 \ \mathrm{eV} $ and mass of $\sim 2 \times
10^{6} \ \mathrm{M_{\sun}} $. The gas distribution is significantly
extended along the minor axis of the galaxy suggesting that it
may be outflowing in the direction perpendicular to the galactic disk.
The mass and energy supply from evolved stars and type Ia
supernovae is sufficient to sustain the outflow.
We also detect a shadow cast on the gas emission by spiral arms
and the 10-kpc star-forming ring, confirming significant extent of the
gas in the ``vertical'' direction.
(iii) Hard extended emission
from spiral arms, most likely associated with young stellar objects
and young stars located in the star-forming regions. The $
\mathrm{L_{X}/SFR}$ ratio equals $\sim 9 \times 10^{38} \
\mathrm{({erg/s})/(M_{\sun}/\mathrm{yr})}$ which is about $\sim 1/3 $
of the HMXBs contribution, determined earlier from Chandra observations
of other nearby galaxies. 
\end{abstract}


\section{Introduction}

The X-ray radiation from the majority of galaxies is
dominated by X-ray binaries  \citep[e.g.][]{fabbiano}. 
In addition, extended emission is present
in galaxies of all morphological types.   
At least part of this  emission is associated with stellar
population and is a superposition of a large number of faint compact
sources -- accreting white dwarfs, active binaries and other types of
stellar sources, old and young \citep{ridge}. 
There is also a truly diffuse component -- emission from ionized gas of 
sub-keV temperature. Its importance varies from galaxy to galaxy, with 
luminous gas-rich ellipticals, like NGC 1316  \citep{kim} and dwarf
galaxies similar to M32  \citep{revnivtsev1}
representing  the two opposite ends of the range. Significant diffuse
or quasi-diffuse emission is also associated with star-formation, the 
Antennae \citep[e.g.][]{baldi} being one of the nearby
examples. The morphology of the gas in starburst galaxies often 
indicates outflows, driven by the energy input into ISM from core
collapse supernovae. Theoretical considerations suggest that gas 
in low mass elliptical galaxies may also be in the state of 
outflow \citep{david}. The mass and energy budget of the ISM in 
this case is maintained by winds from evolved stars and Type Ia 
supernovae. The overall X-ray radiation from a galaxy is a 
superposition of these (and possibly other) components, 
their relative importance being defined by the morphological  
type of the galaxy and its star-formation and merger history. 

Our close-by neighbor,  M31 galaxy gives a unique opportunity to
explore a ``full-size'' spiral galaxy similar to the Milky Way without 
complications brought about by projection and absorption effects,
often hampering studies of our own Galaxy.
Not surprisingly, it has been extensively investigated by every major
observatory of the past decades. Observations by the {\em Einstein}
observatory demonstrated that X-ray binaries account for the most of
the X-ray emission from the galaxy \citep{einstein}. Using the
complete set of  {\em Einstein} data \citet{trinchieri} constrained 
possible amount of ionized gas in the bulge of the galaxy by 
$\la 2 \times 10^{6}  \mathrm{M_{\sun}}$. Based on the {\em ROSAT} 
observations,  \citet{primini} found evidence for 
extended emission component with luminosity of $\sim 6\times 10^{38}$
erg/s in the 0.2--4 keV band. They suggested, that this emission may be of
truly diffuse origin or due to a new class of X-ray sources.  
The unresolved emission from M31 was investigated further by 
\citet{supper}, \citet{west}, \citet{irwin1}, \citet{borozdin}. In all these
studies the existence of a soft emission component with temperature
$kT\sim 0.3-0.4$ keV has been confirmed, although different authors
suggested different explanations of its origin. With advent of Chandra
and XMM-Newton, the consensus seemed to be achieved in favor of truly 
diffuse origin of the soft emission component \citep{chanxmm}.
However, recent progress in understanding the nature of the Galactic
Ridge emission as a superposition of a large number of faint stellar
type sources \citep{revnivtsev2}, made it worth to revisit
the problem of the origin of  extended emission in M31.  
Also, with more Chandra and XMM-Newton observations, more accurate and
detailed investigations became possible. 
Recently \citep{li} analyzed large Chandra dataset of M31
bulge observations and demonstrated presence of both ionized gas and
emission of faint compact sources associated with old stellar
population. Moreover, they revealed  peculiar morphology of the gas
emission and suggested that the X-ray gas in the bulge of M31 may be in
the state of outflow.

\begin{figure}
\includegraphics[width=8.5cm]{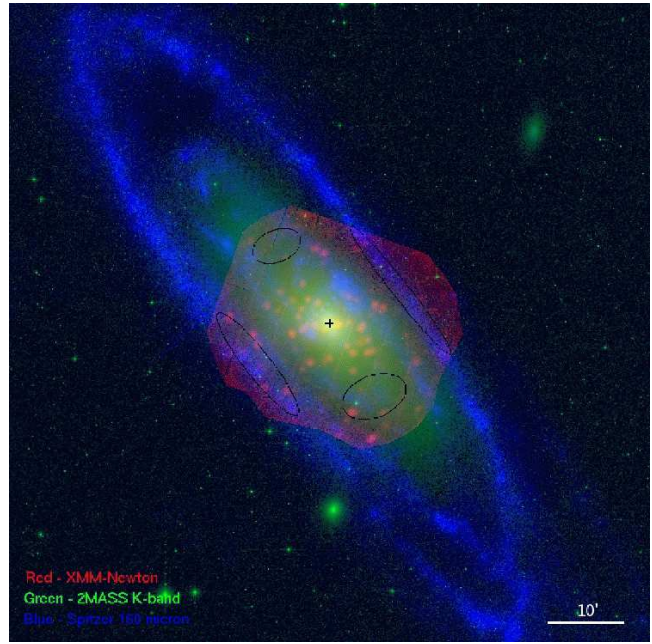}
\caption{RGB image of M31. The colors are as follows:
red is the XMM-Newton data in the $ 0.5 - 1.2 \ \mathrm{keV} $ band,
green is an image of the K-band light from 2MASS and blue is the  $
160 \ \mathrm{\mu m} $ image of Spitzer. Center of the M31 is marked
with a cross. North is up and east is left.} 
\label{fig:rgb}
\end{figure}

In the present paper we combine extensive set of
Chandra and XMM-Newton observations to obtain a broad band and large
field view of the X-ray emission originating in and around the bulge
of M31. We restrict our study to the central region of
$\sim 20\arcmin$ in radius, covered well by Chandra and XMM-Newton
observations currently available in the public archives of these
observatories (Fig.\ref{fig:rgb}). The investigated region has the linear size of 
$\sim 4$ kpc along the major axis of the galaxy, but extends out to  
$\sim 16$ kpc along the plane of the galaxy in the minor axis
direction, due to rather  large inclination angle of M31, 
$i\sim 77\degr$  \citep{henderson}. 
We assume the distance to Andromeda of $780 \ \mathrm{kpc} $ 
\citep{distance}. The Galactic absorption towards M31 is 
$ 6.7 \times 10^{20} \ \mathrm{cm^{-2}} $ \citep{dickey}. 

The paper is structured as follows. In Section 2 we describe the data
and its reduction. We introduce our results in Section 3, where we
present the spatial distribution, morphology and the spectra of the
extended emission. In Section 4 the origin, properties and physical
parameters of different components are discussed and in Section 5 we
summarize our results.

\section{Data reduction}

\indent
We combine data from Chandra and XMM-Newton satellites
adding their benefits together. The primal advantage of Chandra is its  
good angular resolution which allows us to resolve individual X-ray
binaries everywhere including the very central region of the
bulge. XMM-Newton provided better coverage of M31 and collected more
photons, thanks to its larger effective area. It is more
suitable to study the outer part of M31. On the other hand the higher
and less predictable background of XMM-Newton complicates study of low
surface brightness regions. 

\subsection{Chandra}
\label{sec:reduction_chandra}

We used $ 21 $ archival Chandra observations listed in Table 1, taken
between 13.10.1999 and 23.05.2004. For the analysis we extracted
data of the ACIS-I array except for OBS-ID 1575 where we used only the
S3 chip.  
The pattern of available Chandra observations allows us to study the central 
$\approx 15\arcmin$ region. The data reduction was performed using
standard CIAO\,\footnote[1]{http://cxc.harvard.edu/ciao/} software
package tools (CIAO version 3.4; CALDB version 3.4.1).  
For each observations we filtered out the flare contaminated
intervals, excluding the time intervals where the count rate deviated
by more than $20\%$ from the mean value. The resulting
effective exposure times are given in Table 1.

Crucial for the
analysis of the low surface brightness outer regions
is the accuracy of the background subtraction. In treating the
Chandra background we generally followed the procedures
outlined in \citet{hickox}.
We determined the level of the instrumental background using the
stowed data 
set\footnote[2]{http://cxc.harvard.edu/contrib/maxim/stowed/}. In the  
stowed position of ACIS detectors the sky emission is blocked and only
the instrumental background gives a contribution. As demonstrated in the
above-mentioned paper, although the instrumental background level varies
with time, its spectrum remains unchanged. 
The effective area of Chandra is negligible above 9 keV
and the count rate is dominated by the instrumental background. 
Therefore the  $ 9.5 - 12 \ \mathrm{keV}$ count rates can be used to
renormalize background spectra obtained from the stowed datasets.
In dealing with cosmic background we took into account that it
consists of the soft emission associated with the Galaxy and harder
extragalactic component and treated them separately. For the soft
Galactic component we used the best fit spectrum from 
\citet{hickox}. For the extragalactic background
we took into account that some fraction of it has been resolved in our
data analysis procedure and removed along with X-ray binaries.
Using the incompleteness function from \citet{voss}, that was 
obtained using essentially the same data set, we estimated 
that our point source detection sensitivity in the outer 
regions is  $\approx 5 \times 10^{35} \ \mathrm{erg/s}$. We used
the sensitivity for outermost regions because the CXB subtraction
plays role only in these regions where the surface brightness of the
source emission is low. This point source sensitivity results in  
the resolved CXB fraction of $\approx 50\%$, according to \citet{moretti}.
The Galactic and cosmic backgrounds were subtracted from the
vignetting corrected images and profiles. 

\begin{table}
\caption{The list of Chandra observations used for the analysis.}
\centering
\begin{tabular}{c|c|c|c|c}
\hline
Obs ID & $ T_{\mathrm{original}} $ &  $ T_{\mathrm{eff}} $ & Instrument & Date  \\
\hline
$ 303 $  & $ 12.0 $ ks &$ 8.2 $ ks  & ACIS-I & 1999 Oct 13 \\
$ 305 $  & $ 4.2 $ ks & $ 4.0 $ ks   & ACIS-I & 1999 Dec 11 \\   
$ 306 $  & $ 4.2 $ ks & $ 4.1 $ ks   & ACIS-I & 1999 Dec 27 \\   
$ 307 $  & $ 4.2 $ ks & $ 3.1 $ ks   & ACIS-I & 2000 Jan 29 \\   
$ 308 $  & $ 4.1 $ ks & $ 3.7 $ ks   & ACIS-I & 2000 Feb 16 \\
$ 311 $  & $ 5.0 $ ks & $ 3.9 $ ks   & ACIS-I & 2000 Jul 29 \\   
$ 312 $  & $ 4.7 $ ks & $ 3.8 $ ks   & ACIS-I & 2000 Aug 27 \\   
$ 1575 $ & $ 38.2 $ ks& $38.2 $ ks & ACIS-S & 2001 Oct 05 \\
$ 1577 $ & $ 5.0 $ ks & $ 4.9 $ ks   & ACIS-I & 2001 Aug 31 \\
$ 1583 $ & $ 5.0 $ ks & $ 4.1 $ ks   & ACIS-I & 2001 Jun 10 \\
$ 1585 $ & $ 5.0 $ ks & $ 4.1 $ ks   & ACIS-I & 2001 Nov 19 \\   
$ 2895 $ & $ 4.9 $ ks & $ 3.2 $ ks   & ACIS-I & 2001 Dec 07 \\   
$ 2896 $ & $ 5.0 $ ks & $ 3.7 $ ks   & ACIS-I & 2002 Feb 06 \\
$ 2897 $ & $ 5.0 $ ks & $ 4.1 $ ks   & ACIS-I & 2002 Jan 08 \\
$ 2898 $ & $ 5.0 $ ks & $ 3.2 $ ks   & ACIS-I & 2002 Jun 02 \\
$ 4360 $ & $ 5.0 $ ks & $ 3.4 $ ks   & ACIS-I & 2002 Aug 11 \\   
$ 4678 $ & $ 4.9 $ ks & $ 2.7 $ ks   & ACIS-I & 2003 Nov 09 \\    
$ 4679 $ & $ 4.8 $ ks & $ 2.7 $ ks   & ACIS-I & 2003 Nov 26 \\
$ 4680 $ & $ 5.2 $ ks & $ 3.2 $ ks   & ACIS-I & 2003 Dec 27 \\
$ 4681 $ & $ 5.1 $ ks & $ 3.3 $ ks   & ACIS-I & 2004 Jan 31 \\   
$ 4682 $ & $ 4.9 $ ks & $ 1.2 $ ks   & ACIS-I & 2004 May 23 \\
\hline                                       
\end{tabular}                                
\end{table}

In order to study the diffuse emission we need to exclude
contribution of bright LMXBs. According to the luminosity functions of
LMXBs and faint compact sources associated with old population
\citep{lmxb}, the contribution of the former
is defined by the sources more luminous than $\log L_X\sim 35.5-36.0$.
Below this  threshold active binaries and cataclysmic variables are
the dominating X-ray sources. The individual Chandra observations are
too short (typically, $\sim 4$ ksec, see Table 1) to achieve this
sensitivity, therefore we ran point source detection on the combined
image with total exposure of $ T_{\mathrm{eff}} = 112.6 
\ \mathrm{ks} $. To combine the data, each observation was
projected into the coordinate system of OBS-ID 303 and the
attitude corrections from \citet{voss} were applied
in order to better co-align individual event lists. To detect point
sources we ran CIAO \texttt{wavdetect} tool in the $ 0.5 - 8 \
\mathrm{keV} $ band. Some parameters were changed from the default
values to fit our aims. The scales on which 
we were looking for sources were the $ \sqrt{2} $-series from $ 1.0 $
to $ 8.0 $. To minimize the contribution of residual counts from point
sources to diffuse emission we increased the value of the sigma
parameter to $ 4 $; this parameters describes the size of elliptical
source detection regions in standard deviations assuming a 2D Gaussian
distribution of source counts. With these adjustment
we obtained larger source regions than usual, including larger fraction 
of source counts.
To increase sensitivity  we did not filter out flare containing time
intervals for point source detection. The resulting source list 
consisted totally of $ 238 $ sources in the investigated area; the
detected sources are in good agreement with results of \citet{voss}. The 
sensitivity limit in the central region was $ 10^{35} \
\mathrm{erg/s} $, while in the outermost region it deteriorated to
$ 5\times 10^{35} \ \mathrm{erg/s} $. 
Extracting the point spread function with \texttt{mkpsf} for each
source we calculated the fraction of counts inside the source
cell. For most of the sources this fraction exceeded $ 98 \% $,
if it was smaller then the source cells were enlarged accordingly. The
output source cells were used to mask out the point sources 
for further image and spectral analysis.

We produced exposure maps using a two component spectral
model consisting of an optically-thin thermal plasma emission with
temperature of $ 0.30 \ \mathrm{keV} $ and a powerlaw component with a
slope of $ \Gamma = 1.80 $. The ratio of the normalizations of these
two components was $3/1$. This is the best fit two component  
model to the spectrum of the central $ 200 \arcsec $ region. 

\subsection{XMM-Newton}
\begin{table}
\caption{The list of XMM-Newton observations used for the analysis.}
\centering
\begin{tabular}{c|c|c|c}
\hline
Obs ID & $ T_{\mathrm{original}} $ &  $ T_{\mathrm{eff}} $ &  Date  \\
\hline
$ 0109270101 $  & $ 62.5 $ ks & $ 16.0 $ ks  & 2001 Jun 29  \\
$ 0112570101 $  & $ 61.1 $ ks & $ 52.6 $ ks  & 2002 Jan  6  \\   
$ 0112570401 $  & $ 31.0 $ ks & $ 25.0 $ ks  & 2000 Jun 25  \\   
$ 0202230201 $  & $ 18.3 $ ks & $ 17.8 $ ks  & 2004 Jul 16  \\   
$ 0202230401 $  & $ 14.6 $ ks & $ 9.0 $ ks   & 2004 Jul 18  \\
$ 0202230501 $  & $ 21.8 $ ks & $ 2.0 $ ks   & 2004 Jul 19  \\   
\hline                                       
\end{tabular}                                
\end{table}

We analyzed $ 6 $ archival XMM-Newton observations taken between
25.06.2000 and 19.07.2004, as listed in Table 2. We used the
data of the European Photon Imaging Camera (EPIC) instruments \citep{epic}. 
For data reduction we used Science Analysis System (SAS) 
version 7.1. 

In order to exclude the flare contaminated time intervals we double
filtered the lightcurves using hard-band ($ E > 10 \ \mathrm{keV}  $)
and soft-band $ (E = 1-5 \ \mathrm{keV} $) energy ranges according to
\citet{nevalainen}, using $ 20\% $ threshold for
deviation from the mean count rate.  
The remaining useful exposure time is about $ 58
\% $ of the original value. The data was cleaned from the out of time
events using the Oot event lists. 

The observations were re-projected into the coordinate system of OBS-ID
0112570101 and merged together. For point source removal we combined the 
source list obtained from Chandra and XMM-Newton observations. 
In regions which lied outside the field of view of Chandra 
we ran the SAS source detection task to complement the 
Chandra source list. The source regions were enlarged 
to account for larger size of the point spread function of 
XMM-Newton mirrors, their size was adjusted to approximately match
$ \sim 90-98 \% $ PSF encircled energy radius 
depending on the brightness of the point source.
It was not possible to reliably exclude point source 
contribution in the crowded  central $\sim 100 \arcsec$ region, 
therefore it was not used in the following analysis.
Exposure maps were calculated using the \texttt{eexpmap} command of
SAS. In transforming the counts to flux units we assumed the same
spectrum as for Chandra. 

The particle background on EPIC CCDs consists of two components. 
The ``internal'' component is generated in interactions of cosmic rays  
with the detector material and is approximately uniform
across the detector. The second component is due to low energy solar
protons, concentrated by the mirror systems of the telescopes; 
it is vignetted by the mirrors response but the vignetting is flatter
than that for photons.  
The level of both background components is variable
(see http://www.star.le.ac.uk/$\sim$amr30/BG/BGTable.html for details).
According to this we performed the background subtraction in two
steps. At the first step the corners of the CCDs which lie outside of
the field of view, were used to determine the level of the flat
internal background. The obtained background level was subtracted from
each observations individually. The combined contribution of the 
solar protons component and cosmic background was approximately
determined from the observations of nearby fields without extended
sources and subtracted from the final vignetting corrected image. This
method is not perfect, due to the difference in the shape of the
vignetting function between solar protons and photons and due to
variability of the solar protons level.
However, good agreement between emission spectra obtained
from Chandra and XMM-Newton data confirms adequate accuracy of this
procedure.

\section{Results}
\subsection{Images}

The RGB image in Fig.1 presents the XMM-Newton data (red)
overlaid on the $160 \ \mu m$ Spitzer image \citep{rieke} (blue) 
and K-band image from 2MASS Large Galaxy Atlas \citep{jarrett} (green). 
Although the main purpose of this image 
is to show the X-ray data coverage, it crudely illustrates 
the presence of a large population of compact sources as 
well as of the extended emission. It also demonstrates 
the effect of the spiral arms  on X-ray surface brightness distribution. 

The brightness distribution of the extended emission, after
removal of the point sources, is shown in Fig. 2 along with the
contours of the K-band brightness.
The X-ray image was constructed from Chandra data in the $ 0.5 - 1.2 \
\mathrm{keV} $ band. The point sources were excluded and their
locations were filled up with the average local background around
the sources. The X-ray image is adaptively smoothed. In order to 
compare the X-ray surface brightness with the distribution of the
stellar mass we also show K-band contours. The
surface brightness of the extended emission approximately follows
the K-band distribution but the image suggests that the agreement is
not perfect and some distortions in the east-west direction may be
present. 
In order to investigate this in details we consider profiles along the
major and minor axes of the galaxy.  

\begin{figure}
\includegraphics[width=8.5cm]{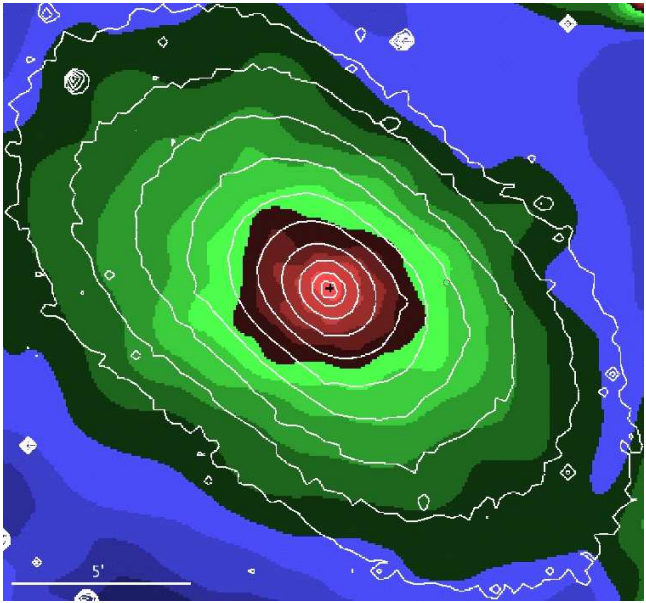}
\caption{Adaptively smoothed Chandra image in the $ 0.5 - 1.2 \
\mathrm{keV} $ band overplotted with K-band contours. The point
sources were removed and their locations were filled with the local
background value. Center of the M31 is marked with a cross. North is 
up and east is left.}  
\end{figure}

\subsection{Surface brightness distribution along the major and minor
axes} 
\label{sec:prof}

\begin{figure}
\resizebox{\hsize}{!}{\includegraphics{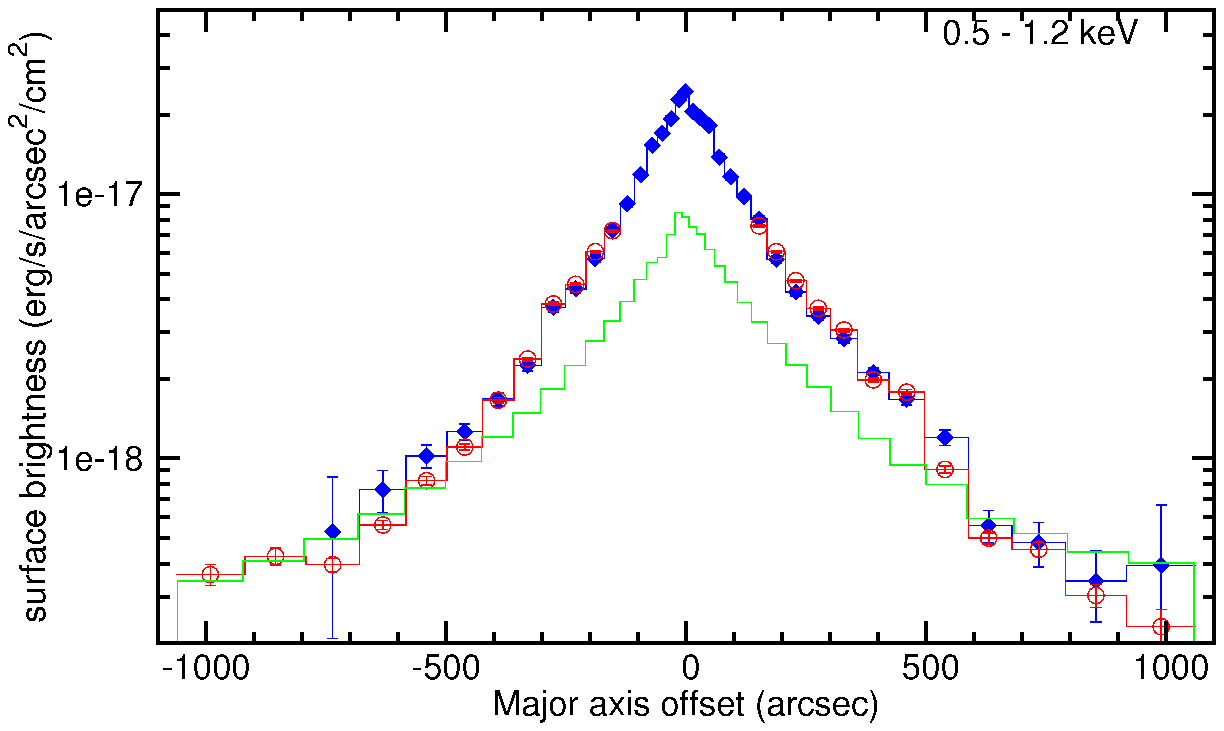}}
\resizebox{\hsize}{!}{\includegraphics{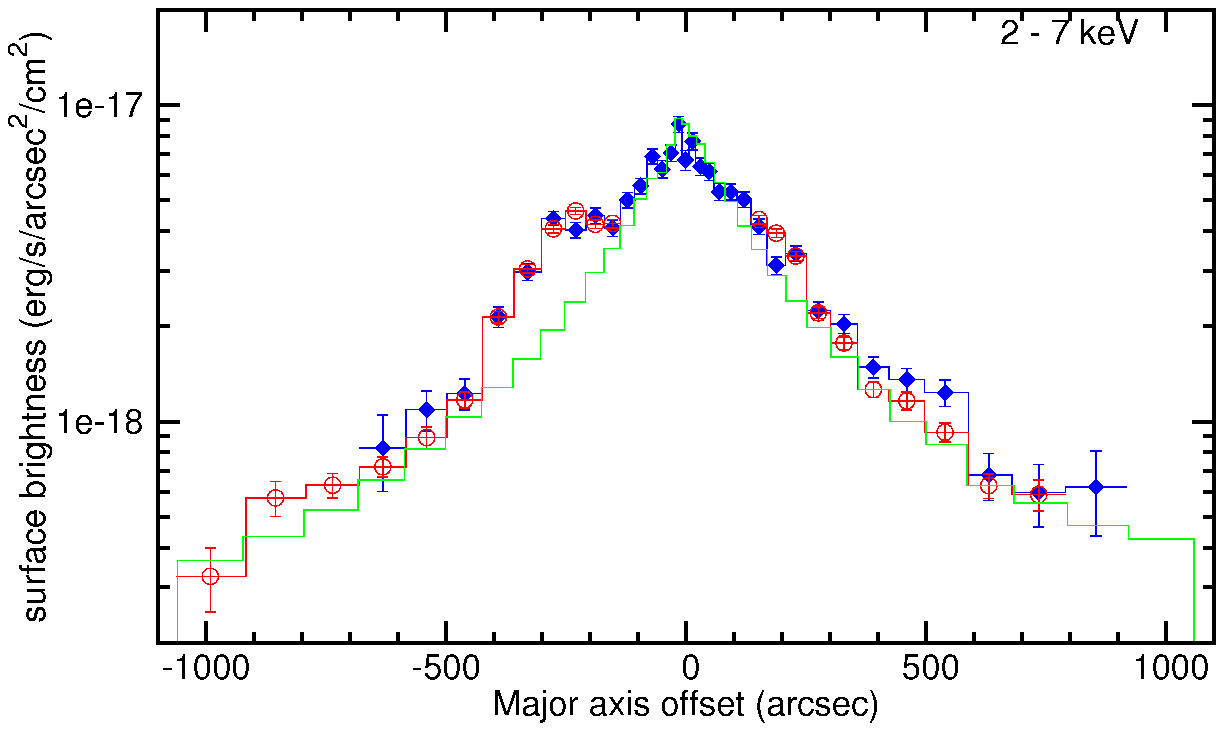}}
\caption{Surface brightness distribution along the major axis. 
in the $0.5 - 1.2 \ \mathrm{keV}$ (upper panel) 
and  $ 2 - 7 \ \mathrm{keV}$ (lower panel) energy bands, background
subtracted. 
The filled symbols (blue in the color version of the plot) show the
Chandra data, open symbols (red) show the XMM-Newton data and the
solid histogram (green) without symbols is the normalized K-band
brightness. The normalization factors  are 
$4\times 10^{27}$ and $ 3 \times 10^{27} \ \mathrm{erg \ s^{-1 } \
L_{\sun}^{-1}} $ for soft and hard band respectively. 
The x-coordinate increases from south-west to north-east. 
}  
\label{fig:prof_major}
\end{figure}

We  studied brightness distribution in two energy bands, $ 0.5 -1.2 \
\mathrm{keV} $ and $ 2 - 7 \ \mathrm{keV} $. Our choice had been
motivated by the shape of the spectrum of extended emission discussed
in the next section. The
profiles were constructed along the major and minor axis with position
angles of $ 45 \degr $ and $ 135 \degr $ respectively. For each profile
surface brightness was averaged over $500\arcsec$ in the transverse
direction, corrected for vignetting and the estimated background level
was subtracted. For XMM-Newton the background level was  
adjusted to achieve better agreement with Chandra  profiles. The
adjusted values were well within the range of the background levels
observed in individual blank-sky observations and differed from the
average blank-sky level by $\la 20\%$.  
The
values of  background were $(4.3,\ 5.6,\ 5.2,\ 6.2)\times 10^{-19}
\ \mathrm{erg\ s^{-1} \ arcsec^{-2} \ cm^{-2}}$
for Chandra (soft, hard) and XMM-Newton (soft, hard) respectively.
For all distributions we found good agreement between the Chandra and
XMM-Newton data. 
We compare X-ray distribution with profiles of the K-band emission. As it
is well-known, the latter is a good 
stellar mass tracer. The K-band profile was obtained for the same
regions as the X-ray profiles, in particular the same source regions 
were excluded in computing all profiles. The normalizations
for the K-band profile are 
$4\times 10^{27}$ and $ 3 \times 10^{27} \ \mathrm{erg \ s^{-1 } \
L_{\sun}^{-1}} $ for soft and hard band respectively.

On Fig.\ref{fig:prof_major} we show the distribution along the major
axis.
In the $0.5 - 1.2 \ \mathrm{keV}$ band the profile shows an excess emission
in the central part of the bulge. At bigger central distances the
X-ray and K-band light follow each other. In the hard band the X-ray
surface brightness follows the near-infrared light
distribution rather well at all central distances, with exception of
the shoulder at the offset of $-300\arcsec$. 
The excess luminosity
of the shoulder above the level suggested by the K-band profile is
$\sim 2\times 10^{37}$ erg/s.  
Its origin is unclear. There is no any easily identifiable feature in
the  image with which it could be associated. 
It can not be due to residual contamination from point
sources. Indeed, the  excess count rate associated with the 
shoulder is $\sim 20\%$ of the total count rate of all point sources
detected in this  region. This is much larger than the expected
residual contamination from point sources in Chandra images, $\la
2-3\%$. Good agreement 
between Chandra and XMM-Newton data also excludes the possibility that
it is caused by an irregularity in the instrumental background.

\begin{figure}
\resizebox{\hsize}{!}{\includegraphics{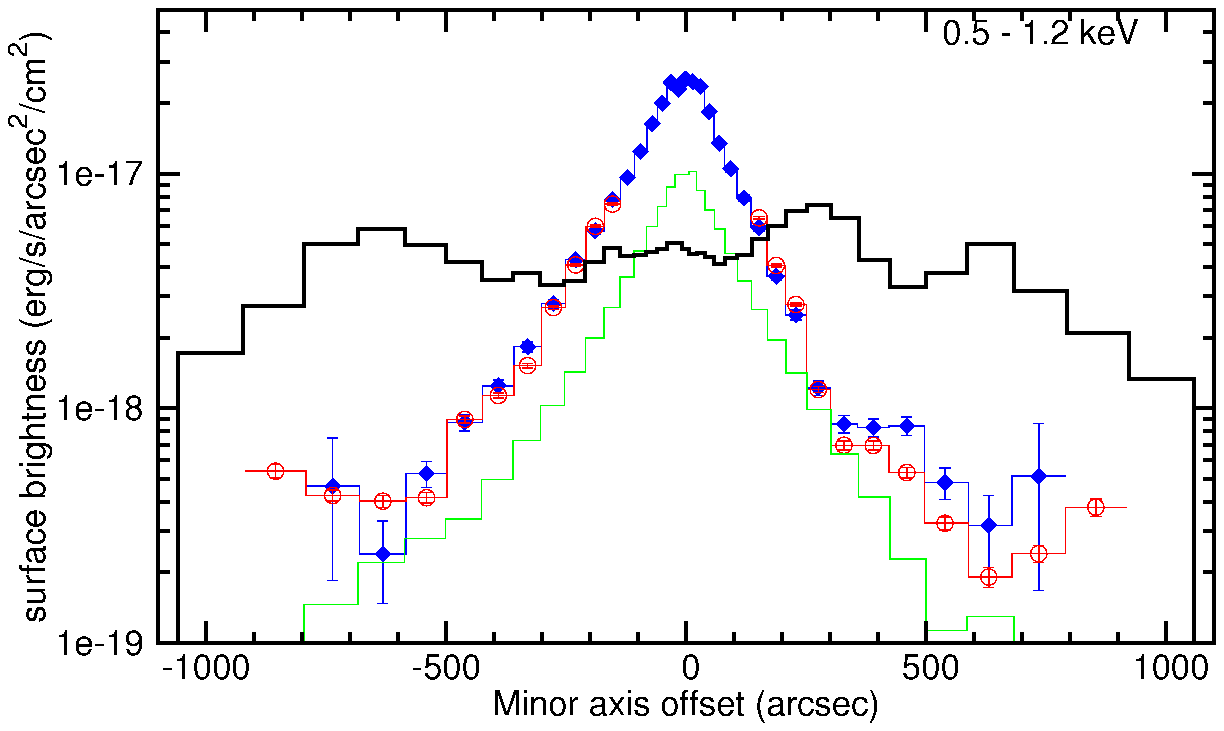}}
\resizebox{\hsize}{!}{\includegraphics{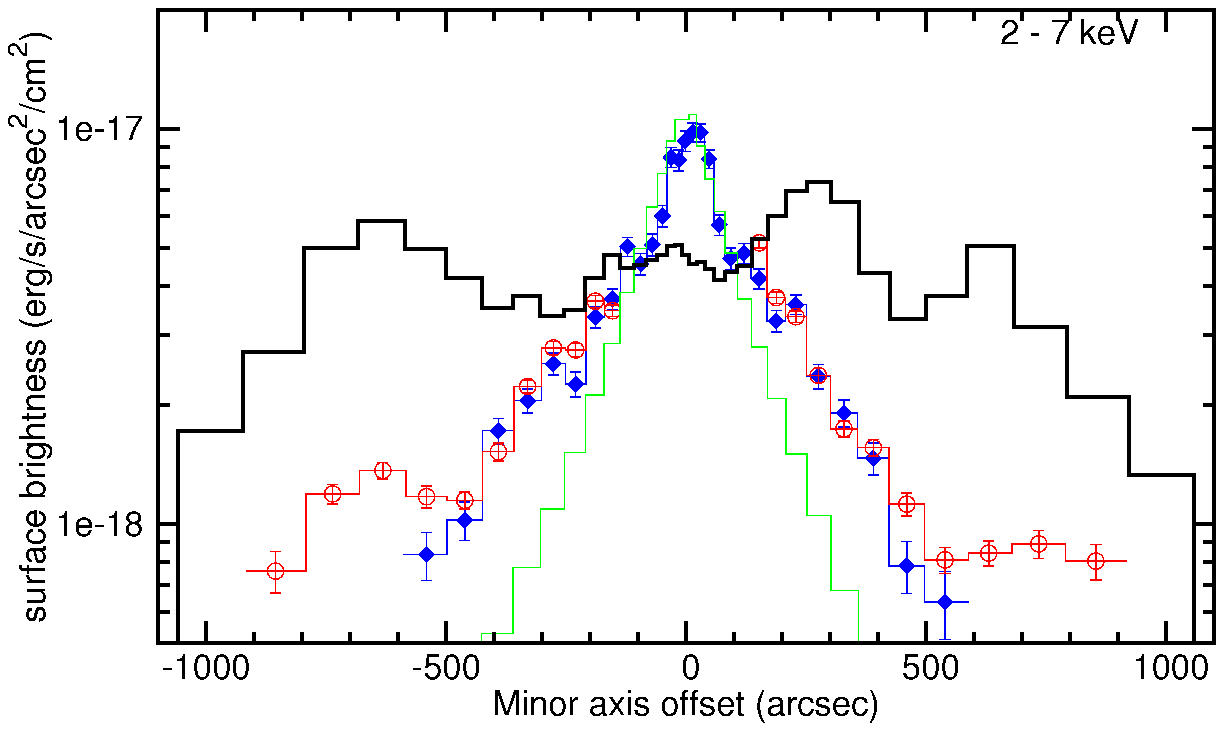}}
\caption{Same as Fig.\ref{fig:prof_major} but along the minor
axis. The normalizations of the K-band profile are same as in
Fig.\ref{fig:prof_major}. The solid histogram with multiple peaks
(black) shows the distribution of the  $160 \ \mathrm{\mu m}$ emission
as obtained by Spitzer.
The x-coordinate increases from south-east to north-west. 
}  
\label{fig:prof_minor}
\end{figure}

The soft band profile along the minor axis (Fig.\ref{fig:prof_minor})
at all offsets exceeds the level suggested by the K-band profile
normalized according to the X/K ratio from the major axis
distribution. Moreover, the  X/K ratio increases
significantly outside $\sim 500\arcsec$ from the center -- the X-ray
distribution appears to have ``wings'' extending out to $\sim
900\arcsec$ or more.  Note that exact shape of the surface
brightness distribution at large offsets from the center
depends on the adopted blank-sky level. The latter can not be
directly determined from the currently available data, due to its
limited field of view.  For this reason we used the average CXB level
and corrected it for the fraction of resolved background sources, as 
described in section \ref{sec:reduction_chandra}. The obtained value,
$\sim (4-5)\times 10^{-19}$ erg/s/cm$^2$, is comparable to 
the remaining (background subtracted) flux as can be seen in
Fig.\ref{fig:prof_minor}. Therefore the extend of the X-ray emission
at large off-center angles can not be unambiguously constrained from
the present data. 
In order to eliminate this uncertainty, more extensive Chandra
observations, including large offset angles are needed.
We note however, that the existence of unresolved emission at large
offsets  is independently confirmed by the east-west asymmetry of the
shadow cast by the 10-kpc star-forming ring, as discussed below.

\begin{figure}
\resizebox{\hsize}{!}{\includegraphics{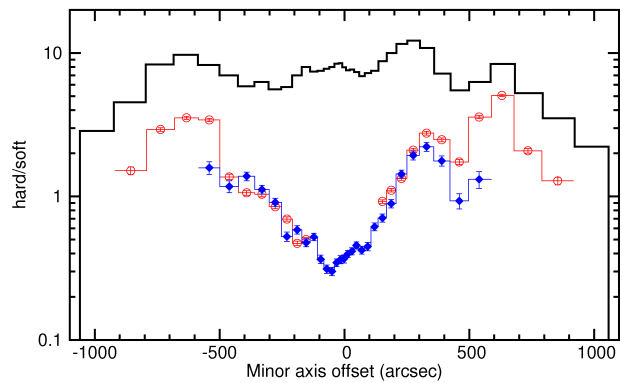}}
\caption{The X-ray hardness ratio profile along the minor axis,
computed using the energy bands  from 
Fig.\ref{fig:prof_minor}. The Chandra and XMM-Newton data are
presented by the same symbols as in Fig.\ref{fig:prof_minor}. The
solid histogram (black) shows the distribution of the  $160 \
\mathrm{\mu m}$ emission  obtained by Spitzer.
The x-coordinate increases from south-east to north-west.}
\label{fig:prof_hr}
\end{figure}

There is a clear asymmetry between eastern and western halves of the
profile, the latter being notably suppressed at offset values of
$+300\arcsec$ and $+600\arcsec$.
The origin of this asymmetry becomes clear after comparison with the 
$160\ \mathrm{\mu m}$ Spitzer profile. 
The far-infrared profile, plotted in Fig.\ref{fig:prof_minor} shows 
several prominent peaks, corresponding to the 10-kpc star-forming
ring and inner spiral arms \citep{gordon}. 
The surface brightness suppressions in the western side of the profile
(positive offsets) correspond to the spiral arm and the 10-kpc
ring. On the other hand no significant effect of the 10-kpc ring can
be seen on the eastern side of the galaxy (negative offsets).

The spatial
orientation of M31 plays an important role in understanding the
link between the X-ray and far-infrared distributions. We see the galaxy
with approximately  $ 77 \degr $ inclination 
\citep{henderson} and the western side of the galactic disk is closer
to us \citep{simien}. The effect of this is that we see
the western side of the bulge through the spiral arms, so the neutral
gas and dust in the star forming regions cast a shadow on the
extended emission in the soft band. The eastern side of the disk, on
the contrary,  is located behind the bulge and does not obscure its
emission. We estimated the column density of the obscuring material in
the spiral arms using the observed brightness difference between the
eastern and western sides and obtained $ N_{H} \sim 1 - 3
\times 10^{21} \ \mathrm{cm^{-2}} $. These numbers are compatible with
values derived from CO maps \citep{nieten}.  

In the hard band profile along the minor axis we see correlation
between the X-ray and far-infrared emission -- the X-ray brightness
appears to increase at the positions of spiral arms. This suggests
that spiral arms are sources of harder X-ray emission. To further
illustrate the impact of spiral arms on the 
observed X-ray brightness we plot in Fig. 5 the hardness ratio along
the minor axis together with the $160 \ \mathrm{\mu m} $ distribution.
This plot confirms the presence and significance of the
soft emission in the center. The hardness ratio has clear peaks at the
positions of spiral arms, which are caused by two effects --
obscuration by neutral and molecular gas and dust in the soft band and
enhanced hard emission associated with the spiral arms. 

\bigskip

Based on the X-ray brightness distributions we conclude that there are
at least three different components in the unresolved X-ray emission
from the central region of M31:
\begin{enumerate}
\item broad band component, following the distribution of K-band light
(i.e. of stellar mass),
\item soft emission, localized in central $\sim 500\arcsec$ along the
major axis of the galaxy and extending out to $\sim 900\arcsec$ or
more along the minor axis,
\item harder emission from the spiral arms and 10-kpc star-forming
ring.
\end{enumerate}

\begin{figure}
\resizebox{\hsize}{!}{\includegraphics{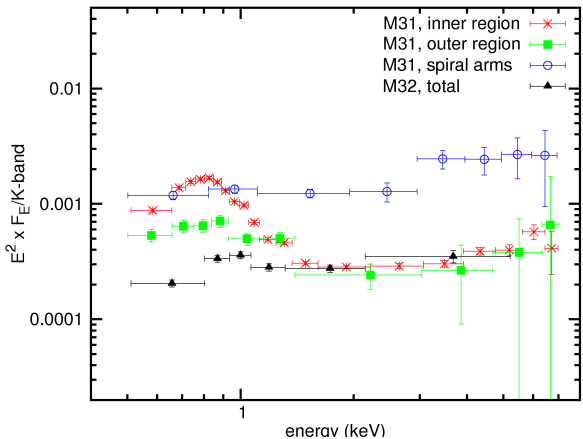}}
\caption{X-ray spectra of different regions in M31 and of M32: 
stars (red) -- inner bulge --  the spectrum of central $200\arcsec$
region; 
filled boxes (green) --  outer bulge  -- combined spectrum of
two elliptical regions at $ \sim 700 \arcsec$ from the center along
the major axis;
open circles (blue) -- spiral arms -- 
two elliptical regions at the 10-kpc ring; 
filled triangles (black) -- spectrum of M32. The M31 regions are shown in
Fig.1 The inner bulge and M32 spectra are obtained by Chandra, outer
bulge and spiral arms -- by XMM-Newton. All spectra are normalized to
the K-band flux.}  
\label{fig:spectra}
\end{figure}

\begin{table*}
\caption{Results of the spectral fits in different regions of M31
using Chandra and XMM-Newton observations. PL denotes the power law
and MKL is the optically-thin thermal plasma emission model.} 
\centering
\begin{tabular}{c||c|c||c|c||c|c||c|c}
\hline
\hline
 & \multicolumn{2}{c||}{Inner region} & \multicolumn{2}{c||}{Outer region} & \multicolumn{2}{c||}{Spiral arms}  & \multicolumn{2}{c}{M32}\\
\hline
Model & $ \Gamma $/kT  & $ \chi^{2} / \mathrm{(d.o.f.)} $ & $ \Gamma $/kT  & $ \chi^{2} / \mathrm{(d.o.f.)} $ & $ \Gamma $/kT  & $ \chi^{2} / \mathrm{(d.o.f.)} $ & $ \Gamma $/kT & $ \chi^{2} / \mathrm{(d.o.f.)} $ \\
\hline\hline
PL &---&---&---&---& $ 2.00 \pm 0.15 $ & $ 292/192 $  & $ 2.07 \pm 0.15 $ & $ 113/72 $  \\
\hline
PL & $ 1.79 \pm 0.10  $ &  & $ 2.48_{-0.44}^{+0.55} $ & &  &  & $ 1.65 \pm 0.21 $& \\
MKL & $ 0.36 \pm 0.01 $ & $ 312/186 $ & $ 0.31 \pm 0.08 $ & $ 127/101 $ & --- &--- & $ 0.54 \pm 0.15 $ & $ 83/70 $  \\
\hline
\hline
\end{tabular}
\label{tab:spectra}
\end{table*}

\subsection{Spectra}
\label{sec:spectra}

We used the ACIS "blank-sky" files in order to subtract the background
from Chandra spectra. As before, we renormalized the background spectra
using the  $ 9.5 - 12 \ \mathrm{keV} $ band count rates. 
For XMM-Newton we used a combined spectrum of observations of nearby
fields as a background spectrum and renormalized it using the $ 10 -
12 \ \mathrm{keV} $ count rate. 
We found good agreement between Chandra and XMM-Newton spectra in all
investigated regions.

In Fig.6 we show  spectra of different regions in M31.
The spectral extraction regions are depicted in Fig.1. The spectra of
outer bulge and spiral arms are based on XMM-Newton data, other
spectra are from Chandra. 
Also shown is the spectrum of M32.
It was obtained using Chandra observations
OBSID  2017, 2494 and 5690 with exposure time of 
$T_{\mathrm{eff}} = 168.5 \ \mathrm{ks} $. The
spectrum was extracted from elliptical region with the position
angle of $40 \degr$ and with $90 \arcsec$ major and $71 \arcsec$ minor
axes; the data analysis procedures were identical to
M31.  All spectra are normalized to the same level of K-band
brightness. The spectral fitting was done in the $ 0.5 - 7 \
\mathrm{keV} $ band. The element abundances were fixed at Solar value,
the hydrogen column density was fixed at the Galactic value. 
The results of spectral fits are summarized in Table 3.

All spectra shown in Fig.\ref{fig:spectra} have a prominent power law
component with the photon index of approximately $\Gamma\sim 2$ and
a soft component of varying strength. It is strongest in the inner
bulge where it by far dominates X-ray emission below 1.2 keV.
The best-fit temperature of the soft component in the M31 spectra
obtained in the two-component model is in the $\sim 0.3-0.4$ keV
range (Table \ref{tab:spectra}). As can be seen from the table, the
simple two-component model consisting of a power law and 
optically-thin thermal plasma emission spectrum (MEKAL model in XSPEC) does not
adequately describe the spectra, especially the spectrum of the inner
bulge, having the largest number of counts. The deviations are
due to the soft band, pointing at the complex shape of the soft
component. With the second MEKAL component the fit improves 
significantly, the best fit parameters for the inner bulge spectrum
are $kT_{1}\approx 0.2\ \mathrm{keV}  $ and $kT_{2}\approx 0.5\
\mathrm{keV}$ with $\chi^2=227$ for $ 184 $ d.o.f. 
Making the element abundance a free parameter improves
the fit quality of the three-component model further to $\chi^2=205$
for $ 183 $ d.o.f. with the best fit abundance of $0.17$ of the solar
value. We considered several other spectral models with free element
abundance, they all produced sub-solar values in the range of 
$0.1-0.2$. We also tried to vary element abundances individually and
found that the model is most sensitive to Ne, Fe and Ni abundances with the
best fit  achieved when 
the abundances of  Ni is a  free parameter. The
two component (vmekal + power law) model requires the Ni
abundance $\approx 3-4$ times solar value (abundances of other
elements were fixed at solar). On the contrary, the models
with the free Fe abundance give a subsolar best fit values for the
latter, $\sim 0.6$. The fit quality improves further with the
non-equilibrium thermal emission model (vnei model in XSPEC) with the
best fit value of $n_e\,t\sim 5\cdot 10^{11}$ sec/cm$^3$ and similar
dependence on the element abundances as for vmekal model. All these
modifications do not change the best fit temperature significantly. 
However, none of the models achieves acceptable values of $\chi^2$.  
It is unclear, how much weight should be given to these
results, as they can be an artifact of the inadequate spectral model and
insufficient energy resolution of the ACIS-I detector. 
Indeed the emission from the inner bulge has a complex spectrum
composed of several constituents of different temperature and
ionization state, some of which may be out of the collisional  
ionization equilibrium.

The outer bulge spectrum has a less prominent soft component,
approximately by a factor of $ 3 $ weaker than the inner
bulge, but its temperature, $kT\approx 0.3\ \mathrm{keV}$, is
compatible with the inner bulge value. 
The spectrum of M32 is similar, although only a very faint soft
component is present here.  Its temperature, $kT=0.54\pm0.15\
\mathrm{keV}$ may be somewhat higher than in M31.  
All three spectra (inner and outer bulge in M31 and M32) nearly
perfectly match each other above $\sim 1.2$ keV, after normalization 
to the K-band flux. This is a strong argument in favor of their
similar origin.

The emission from spiral arms clearly stands out. 
It does not have any significant soft component and, most importantly,
its normalization (per unit K-band flux) is by a factor of $ 4-10 $ higher
in the hard band than for the other spectra. This difference is
smaller in the soft band, however it is still significantly higher
than the spectrum of the outer region. We describe this spectrum with
a powerlaw model and we obtain $\Gamma = 2.0$. As with other spectra,
the large $\chi^2$ value indicates more complex spectral shape.

\subsection{Morphology of the soft excess emission}

\begin{figure*}
\hbox{
\includegraphics[width=8.75cm]{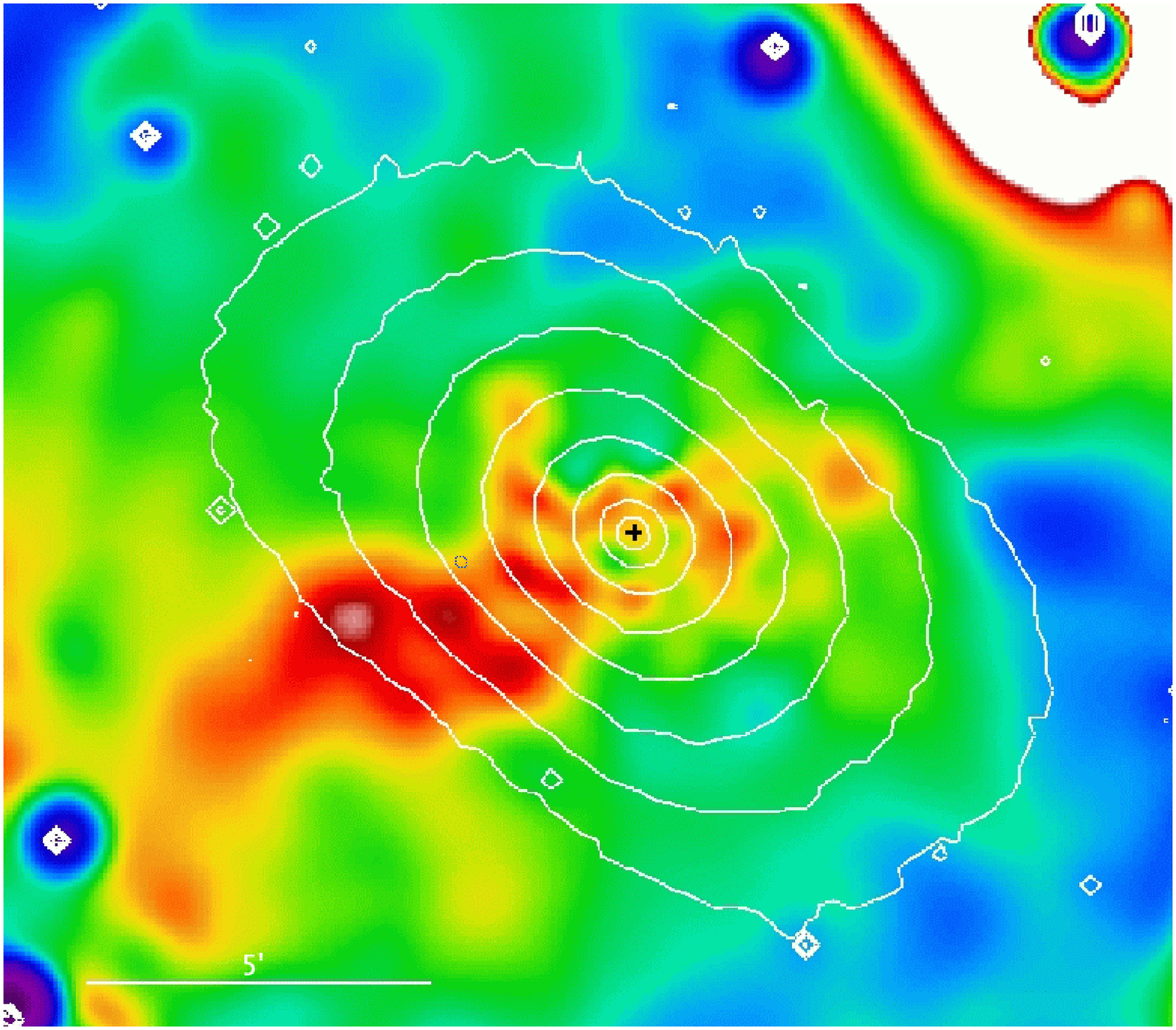}
\hspace{0.25cm}
\includegraphics[width=8.75cm]{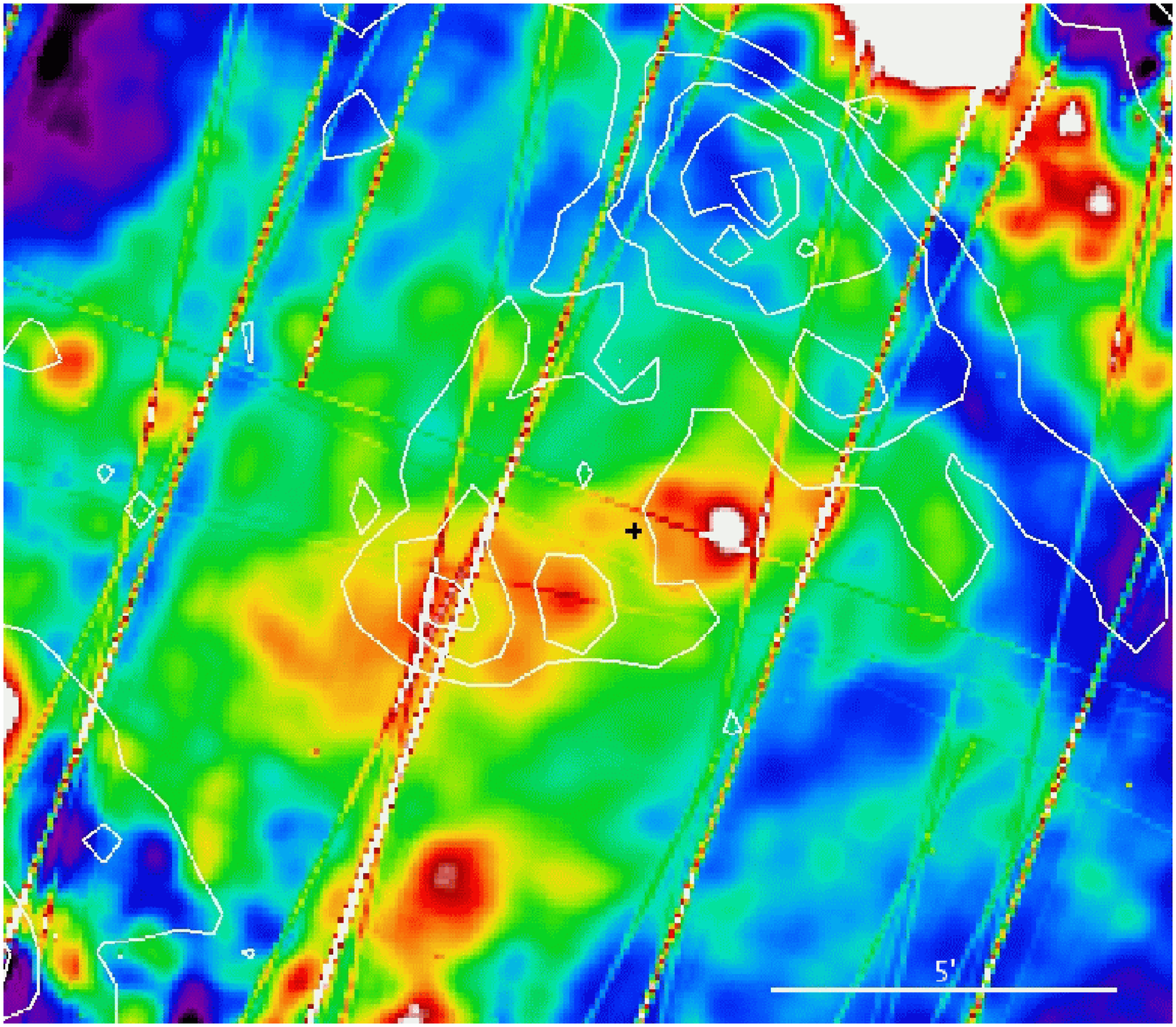}
}
\caption{Ratio of $ 0.5 - 1.2 \ \mathrm{keV} $ image to the the K-band
image obtained from Chandra (left) and XMM-Newton (right) data. After
the removal of the point sources the X-ray images were adaptively
smoothed, the K-band image was smoothed correspondingly. The contours
in the left and right panels show K-band and $160 \ \mathrm{\mu m}$
brightness distributions respectively.   
The center of M31 is marked with a cross. North is up and east
is left.} 
\label{fig:ratio}
\end{figure*}

To explore the spatial distribution of the soft excess emission we use
two approaches. 
Firstly we consider the ratio of the $0.5 - 1.2\ \mathrm{keV}$ X-ray
image to the near-infrared image. The advantage of the ratio image is
that it levels out the component in X-ray emission which is
proportional to the K-band light. The disadvantage is that the X/K
ratio may become very large at large central distances, where the
X-ray emission is unrelated to old stellar population and deviates
significantly from the K-band light distribution.
Before producing the ratio image we excluded point sources 
and filled their locations with local background level using the
\texttt{dmfilth} tool of CIAO. The X-ray images were 
adaptively smoothed and the near-infrared image was convolved with a
gaussian with the width comparable to typical smoothing width near the
center of X-ray image. The X-ray images were exposure corrected, the
particle background and CXB was subtracted.

Fig.\ref{fig:ratio} shows the ratio images obtained from Chandra and
XMM-Newton data. They are consistent with each other and illustrate
very well the 
morphology of the soft excess emission. Comparison with the K-band
contours shown in the left panel demonstrates clearly that the
the soft excess emission has nothing to do with the old stellar
population traced by the near-infrared light. Unlike stellar light,
it is strongly elongated in the approximate direction of the minor
axis of the galaxy extending beyond the boundaries of the bulge and
projecting onto the 10-kpc starforming ring.
The images also reveal strong east-west asymmetry, already seen in the
minor axis distribution (Fig.\ref{fig:prof_minor}).  As discussed in
the section \ref{sec:prof}, this asymmetry is caused by absorption of
soft X-rays by the neutral and molecular gas in the spiral arms on the
north-western side of the galactic disk.
This conclusion is further supported by the anti-correlation between
$160\ \mathrm{\mu m}$ flux and the soft X-ray brightness, with obvious
shadows seen at the positions of spiral arms (Fig.\ref{fig:ratio},
right panel). At the same time, the gas and dust do not have any
significant effect on the soft X-ray brightness in the south-eastern
side of the galaxy.

\begin{figure}
\includegraphics[width=8.5cm]{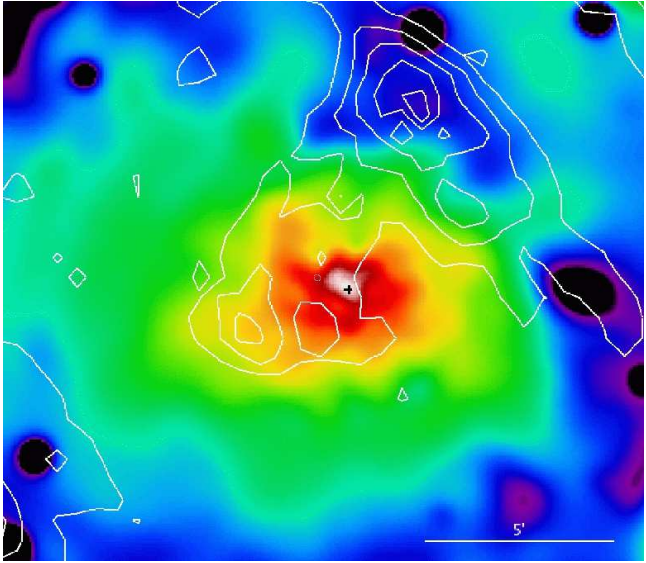}
\caption{The spatial distribution of the soft component. 
The difference between Chandra image in the $0.5 - 1.2\ \mathrm{keV}$
band and the K-band image normalized according to the X/K ratio of the 
outer bulge. The contours show the intensity levels of 
$160\ \mathrm{\mu m}$ Spitzer image. The center of M31 is marked with
a cross. North is up and east is left.} 
\label{fig:residual}
\end{figure}

In the second approach we consider the difference between the X-ray
and the K-band images, the latter renormalized according to the X/K
ratio of the old stellar population component. Unlike the ratio image it
shows the true brightness distribution of the soft excess emission.
For this purpose we used the same $0.5-1.2\ \mathrm{keV}$ Chandra
image as for the ratio image; the normalization of K-band image was $
4 \times 10^{27} \ \mathrm{erg \ s^{-1 } \ L_{\sun}^{-1}} $. 
Fig.\ref{fig:residual} shows the result. The image generally confirms
the overall morphology of the soft emission revealed by the ratio
image. 

The images show significant sub-structures in
the inner $\sim 100\arcsec-150\arcsec$ with the angular scale of
about $30 \arcsec$. The origin of this substructures is not clear,
although some correlation with the position of the peaks on the $160\ 
\mu$m image suggests that at least some of them  may be caused by
absorption. These sub-structures may deserve a special study which is
beyond the scope of this paper.

\subsection{{\boldmath $L_X/L_K$} ratios}

We calculate ratios of X-ray to K-band luminosity for the same
spatial regions as used for the spectral analysis, characterizing  the
inner and outer bulge and spiral arms. 
As suggested by the spectra the  natural boundary between the soft and
hard 
energy bands is $1.2 \ \mathrm{keV} $. However, in order to facilitate 
comparison with previous studies we computed the ratios in the 
$0.5-2\ \mathrm{keV}$ and $2-10 \ \mathrm{keV}$ energy bands. The
X-ray luminosities for these bands were computed using the best fit
spectral models. 
The errors for the luminosity and, correspondingly, 
for X/K ratios account for the model normalization
error only and do not include uncertainties in the
spectral parameters or any other systematic uncertainties. 
In computing the hard band luminosity we extrapolated
the best fit model outside the energy range used for spectral fits,
0.5-7 keV.  The K-band luminosities were calculated in each region
using the 2MASS image. \\ 
\indent
The $L_X/L_K$ ratios are presented in Table 4. 
These numbers can be transformed to X-ray-to-mass ratios dividing them
with the K-band mass-to-light ratio, which is of the order of unity.
Its value can be computed using 
the $ \mathrm{B - V} \approx 0.95 $ colour index from \citet{walterbos1}
and applying the relation between $ M_{\star}/L_{K} $ and 
the B--V colour \citep{bell}, which gives $M_*/L_K\approx 0.85$. 
A close values has been derived by \citet{kent} based on the dynamical
mass measurement, $M_*/L_K\approx 1.1$.  We compare 
the X/K ratios for M31 with M32 and Milky Way. The M32 numbers 
were obtained from the spectra of section \ref{sec:spectra} 
in the same way as for M31. 
For the Milky Way we used results from
\citet{revnivtsev2} and \citet{sazonov}, and transformed them to
$L_X/L_K$ for two values of K-band mass-to-light ratio, 
0.7 \citep{dwek}  and 1 \citep{kent}. 
The X/K ratio in \citet{sazonov} is given for the
$0.1 - 2.4\ \mathrm{keV}$ band, following them we  converted it to the
$ 0.5 - 2 \ \mathrm{keV}$ 
band multiplying  by $0.7$ which is typical ratio for coronally active
stars \citep{fleming}. The contribution of young stars is
excluded. \citet{revnivtsev2} used data from RXTE
in the $ 3 - 20 \ \mathrm{keV} $ and calculated the $ 2 - 10 \
\mathrm{keV} $ luminosities assuming a powerlaw spectrum with $ \Gamma
= 2.1 $. 

\indent
In the $2 - 10 \ \mathrm{keV}$ band the X-ray to K-band ratios for
bulge regions in M31 and for M32 are compatible with each other
and appear to exceed 
slightly the Milky Way value. This may be explained by the difference
in the spectral shape assumed in the flux calculation.
Assuming the same power law with $ \Gamma = 2.1 $ as used by
\citet{revnivtsev2}, we obtained for M31 
$ L_{X}/L_{K} = (3.0 \pm 0.1) \times 10^{27} \ 
\mathrm{erg \ s^{-1 } \ L_{\sun}^{-1}} $, in agreement 
with the Milky Way value. The good agreement of the X-ray to K-band
ratios suggests similar origin of the 2--10 keV emission in all three
galaxies, as discussed in more detail in the following section.

In the soft band, the X/K ratios for M31 are systematically larger
than M32 ones. This is to be expected from the spectra
(Fig.\ref{fig:spectra}) and is due to the presence of the soft excess
emission in M31. 
Also, in interpreting the soft band ratios the interstellar 
absorption should be taken into account. Indeed, \citet{sazonov} studied 
sources in the Solar neighborhood where the absorption is insignificant. 
The M31 and M32 data, on the contrary, is subject to the 
galactic absorption, with values of $6.7 \times 10^{20}$ and $6.3 \times
10^{20}$ cm$^{-2}$ respectively \citep{dickey}. In addition,
M31 has spatially variable internal absorption reaching 
few\,$\times 10^{21}$ cm$^{-2}$ in the spiral arm regions 
\citep{nieten}. The total NH for the outer bulge region in M31 is 
probably in the range $(7-10)\times 10^{20}$ cm$^{-2}$, giving the
absorption corrected value of the soft band X/K ratio 
$ L_{X}/L_{K}\approx(8-10)\times 10^{27}\ \mathrm{erg\ s^{-1 } \
L_{\sun}^{-1}} $. 
For M32 we obtain after absorption correction 
$ L_X/L_K\approx 4.3 \times 10^{27} \ \mathrm{erg \ s^{-1 } \
L_{\sun}^{-1}} $.
Taking into account the uncertainties quoted by \citet{sazonov} and
uncertainties in the K-band mass-to-light ratio
the X/K ratio for Solar neighborhood is formally compatible with the
absorption corrected value for M31 and probably slightly higher than
for M32. 
It is unclear, how much weight should be given to this discrepancy due
to the number and amplitude of uncertainties involved.
The absorption corrected value for the inner bulge of M31 is
definitely 
larger, $L_X/L_K\approx (12-15)\times 10^{27} \ \mathrm{erg \ s^{-1 } \
L_{\sun}^{-1}} $.

The X/K ratios in the spiral arm region are significantly higher.
The additional internal absorption can be upto
few\,$\times 10^{21} \ \mathrm{cm^{-2}}$, resulting in the absorption
correction factor in the soft band as large as $\approx 4$. 
The final X/K ratios in both bands are approximately an
order of magnitude larger than for the Milky Way.

\begin{table}
\centering
\begin{minipage}{8.5cm}
\caption{X-ray to K-band luminosity ratios in different regions in M31
and in M32 compared to the Milky Way. All values are given in units of
$ 10^{27} \ \mathrm{erg \ s^{-1 } \ L_{\sun}^{-1}} $.} 
\begin{tabular}{l||c|c|}
\hline
 &  $ 0.5 - 2 $ keV$^a$ & $ 2 - 10 $ keV  \\ 
\hline
M31 inner bulge & $ 9.4 \pm 0.1 $ & $ 4.5 \pm 0.1 $ \\        
M31 outer bulge & $ 6.2 \pm 0.1 $& $ 3.6 \pm 0.2 $ \\   
M31 spiral arms & $ 22.1 \pm 0.3 $ & $ 31.4 \pm 0.6 $ \\ 
M32 total & $ 3.5 \pm 0.1 $ & $ 4.1 \pm 0.2 $ \\
\hline
Galaxy \citep{sazonov} $^b$ &  $ 7.3-10.5\ (\pm 3) $ & $2.2-3.1\ (\pm 0.8)$ \\  
Galaxy \citep{revnivtsev2} $^b$ & -- & $ 2.2-3.1\ (\pm 0.5) $ \\
\hline
\end{tabular}
$^a$ The X/K ratios for M31 and M32
were not corrected for absorption. See the text for the absorption
corrected values.\\
$^b$ The $X/M_*$ ratios were converted to $X/K$ ratios assuming two
values of the K-band mass-to-light ratio, 0.7 (first number) and 1.0;
the uncertainty given in the parenthesis corresponds to $M_*/L_K=1$
\end{minipage}
\label{tab:xk}
\end{table}

\section{Discussion}

\subsection{Faint compact sources}

We showed that a broad-band emission component exists in M31 which 
(i) follows the K-band light distribution and (ii) its $2-10$ keV X/K 
ratio is identical to that of M32 and Milky Way. 
These suggest beyond reasonable doubt that this component has similar
origin to the Galactic Ridge X-ray emission of the Milky Way
\citep{revnivtsev2}. Namely, it is associated with the old stellar  
population and is a superposition of a large number of weak sources of 
stellar type, the main contributors in the 2--10 keV band being 
cataclysmic  variables and coronally active binaries \citep{sazonov}.

The X/K ratios are compatible between all three galaxies in the 2--10
keV band, but in the soft band they differ in M31 and M32 (Table
4). In the inner bulge this difference is 
clearly due to the 
contribution of the soft emission from the ionized gas, as discussed
below. Although we chose the outer bulge region as far as possible
along the major axis of the galaxy, where the soft X-ray brightness
follows the K-band profile, some residual contribution from the gas
can not be entirely excluded. Therefore it is not clear if this
difference reflects genuine difference between  properties and/or
content of X-ray emitting stellar populations in these two galaxies.    
Due to large systematic uncertainties, the Milky Way value is formally
compatible with both galaxies and any quantitative comparison is
inconclusive at this point.

\subsection{Ionized gas}
\indent
There are several arguments which suggest that the excess soft
component has a non-stellar origin. The most important is the
morphology of the excess emission, namely the striking difference from
the distribution of the near-infrared light (Fig.\ref{fig:ratio}). 
As no significant color gradients are observed in the bulge of M31
\citep{walterbos2}, its stellar content 
must be sufficiently uniform and can not give rise to the observed
non-uniformities in the X/K ratio. Enhanced X-ray to K-band ratio could be
explained if a notable young population was present in the bulge,
which is also not the case \citep{stephens}.
With the stellar origin excluded, it is plausible that the soft excess
emission is of truly diffuse nature and originates from ionized  gas
of $\sim 0.3-0.4$ keV temperature.

To study the  physical properties of the ionized gas we use a
rectangular region on the south-eastern side of the galaxy, to avoid
the complications due to attenuation by the spiral arms on the
north-western side.  
The size of the region is $8\arcmin$ along the major axis and
$10\arcmin$ along the minor axis of the galaxy.
From the spectral fit, the total X-ray luminosity of the soft
component in the $ 0.5 - 2 \
\mathrm{keV} $ energy 
range is $ \approx 10^{38} \ \mathrm{erg/s}$, after the absorption
and bolometric correction the total bolometric luminosity
is $ L_{bol}\approx 2.3 \times 10^{38} \ \mathrm{erg/s} $ assuming
Galactic column 
density. From the emission measure of the gas, $ \int n_{e} n_{H} dV = 6.3
\times 10^{60} \ \mathrm{cm^{-3}}$, we estimate that the mass of the
gas in the studied volume is of the order of $ \sim 10^{6}\
\mathrm{M_{\sun}}$. We note, that assuming the symmetry between the
north-western and south-eastern sides of the galaxy, the total
quantities for the mass and bolometric luminosity are twice the
quoted values.
The average number density is about $n_e \sim 7\times 10^{-3} \
\mathrm{cm^{-3}}$.  The cooling time of the gas is 
$ t_{cool} =  (3kT)/(n_{e} \Lambda(T))\sim 250 $ million years.
We also applied a two temperature model to the soft emission and
obtained for the two components: 
$kT_1\sim 0.25$ keV, $kT_2\sim 0.6$ keV, 
bolometric luminosities of $ \sim 2 \times 10^{38}$ and $ \sim
10^{38}$ erg/s, total masses of $\sim 10^6$ and $\sim 0.6 \times 10^{6}$
M$_{\sun} $ and cooling times of $\sim 200 $ and $\sim 800 $ million years.
We mention that the above computed values strongly depend on 
the applied spectral model.

The morphology of the gas indicates that it is not in the hydrostatic
equilibrium in the gravitational potential of the galaxy. It suggests
rather, that gas is outflowing from the bulge in the direction
perpendicular to the galactic disk \citep[see also][]{li}. 
The mass and energy budget of the
outflow can be maintained by the mass loss from the evolved stars and
Type Ia supernovae.
\citet{knapp} estimated the mass loss rate from
evolved stars for elliptical galaxies 
$ \sim 0.0021 \ L_{K}/L_{K,\sun} \ \mathrm{M_{\sun} \ Gyr^{-1}} $. 
This rate can be applicable to the bulge of M31 as 
the stellar populations are similar. 
The K-band luminosity of this region is $ L_{K} = 1.4 \times 10^{10} \
\mathrm{L_{K,\sun}}$. The estimated total mass loss rate is
$ \approx 0.03 \ \mathrm{M_{\sun}/yr}$. The stellar yields
produce the total amount of the observed gas on a timescale of 
$\sim 35$ million years which is shorter than the cooling
timescale of the gas.

To estimate the energy input from type Ia supernovae we use results of
\citet{mannucci} who give the supernova rate of 
$N_{\mathrm{SN \ Ia}} = 0.035_{-0.011}^{+0.013} \ \mathrm{SNu} $
for E and S0 galaxies, where $ 1 \ \mathrm{SNu} = 1 \ \mathrm{SN}/10^{10}\ 
L_{K,\sun} $ per century. Assuming that one supernova releases 
$ E_{\mathrm{SN \ Ia}} = 10^{51} \
\mathrm{ergs}$ into the interstellar medium, we obtain about $ 1.5
\times 10^{40} \ \mathrm{erg/s} $ energy that goes into ISM. The
minimal energy required to lift the 
gas in the gravitational potential of the galaxy can be calculated
from $ E_{lift}=7.2 \dot{M_{\star}} \sigma_{\star}^{2} $ \citep{david}.
With $ \sigma_{\star} = 156 \pm 23 \ \mathrm{km/s}
$  \citep{lawrie} we obtain $\sim 3.3 \times
10^{39} \ \mathrm{erg/s} $. 
If the gas is heated to the observed temperature by supernovae, it
requires $\sim 10^{39}$ erg/s.
This estimates indicate that
the energy input from supernovae is approximately $\sim 3-4$ times
larger than the minimal energy required to drive a galactic wind from
the galaxy, similar to the result of \citet{david} for low-luminosity 
ellipticals. 

Type Ia supernovae will also contribute
to the chemical enrichment of ISM with iron-peak elements. Typically 
$0.7 \ \mathrm{M_{\sun}} $ of iron 
is provided by each SN Ia event \citep{iron}, that gives about 
$ 3.4 \times 10^{-4} \ \mathrm{M_{\sun}} $ of iron per 
year. Assuming complete mixing of the supernova ejecta with the
stellar wind material we would expect the iron abundance in the hot ISM
$\approx 1.1 \times 10^{-2} $ by mass, which exceeds the Solar value 
of $1.9 \times 10^{-3} $ \citep{anders} by a factor
of $\sim 6$.
The observed spectra  are inconsistent with high iron
abundance in simple one- or two-component thermal models, but these
models can not adequately describe the  spectra anyway, therefore this
result can not be used as a conclusive argument. On the other hand, 
the discrepancy between high predicted and low observed 
abundance of iron is a well-known problem for elliptical
galaxies, where type Ia supernovae also plays important 
role in thermal and chemical evolution of the ISM; it  
has been addressed in a number of studies, e.g. \citet{brighenti}. 
We also note, that although the iron is the primary element by mass in
the  type Ia supernovae ejecta, the ISM will be also enriched by other 
elements, most significantly by nickel. 
The detailed analysis of this problem is beyond the scope of the
present paper.

Assuming that the gas leaves the galaxy in a steady state wind  
along the axis perpendicular to the plane of the disk,
the outflow  speed $ v_{w} $ can be calculated from $\dot{M_{\star}}
= \pi r^{2} \rho_{gas} v_{w} $ where $ \rho_{gas} $ is 
the average gas density estimated above and $r$ is the radius of the
base of the imaginary cylinder filled with the outflowing gas.  
This calculation gives $ v_{w} \sim 60 \ \mathrm{km/s} $
which is smaller by a factor of few than the local sound speed. Such a
slow sub-sonic motion of the gas can not explain the observed
elongated 
shape of the gas distribution, which may be related to the
the magnetic fields and galactic rotation.
We can use the fact that the shadow from the 10-kpc star-forming ring
is present to estimate the  extent of the gas along the axis
perpendicular to the galactic disk. The angular distance of the shadow
from the center is $\sim 600\arcsec-700\arcsec$, giving the
``vertical'' extent of the gas of $\ga 2.5$ kpc.

\subsection{Spiral arms}

The  $\sim 10$ times higher X-ray to K-band ratios observed in the
spiral arms (Table 4) and their different emission spectrum
(Fig. \ref{fig:spectra}) suggest that X-ray emission from spiral arms
has different nature than the bulge. As spiral arms are
associated with star-formation, an obvious candidate is X-ray emission
from young stellar objects (protostars and pre-main sequence stars) 
and young stars, which are well-known sources of X-ray 
radiation \citep{spiral}.  

As X-ray emission from the spiral arms is associated with young
objects, it is natural to characterize it with $\mathrm{L_{X}/SFR}$
ratio. We compute this ratio for the regions used in spectral
analysis. 
The FIR flux was determined from the $ 160\ \mathrm{\mu m} $ Spitzer
image, $ 290 \ \mathrm{Jy}$. In computing this value we
subtracted the blank-sky background of nearby fields.
To convert it to SFR we used results of IR spectral fits
from \citet{gordon}, which gave 
${\rm SFR} = 9.5 \times 10^{-5}\, F_{160\mu}/{\rm Jy}$ M$_{\sun}$/yr 
for the  M31 distance. Thus we obtained the star-formation rate of 
$ 0.028 \ \mathrm{M_{\sun}/yr} $
in the region used for the analysis. 
The X-ray luminosity in the same
region is $ 3 \times 10^{37} \ \mathrm{erg/s} $ in the $ 2-10 \
\mathrm{keV} $ band. After subtracting the X-ray emission due to
the old stellar population we obtain $ 2.6 \times 10^{37} \
\mathrm{erg/s} $ energy. From this we can compute
$ \mathrm{L_{X}/SFR} \approx 9.4 \times 10^{38} \
\mathrm{(erg/s)/(M_{\sun}/yr)} $. This value is $\sim 1/3$ of the total
$ \mathrm{L_{X}/SFR} $ arising from HMXBs which is $ 2.5 \times
10^{39} \ \mathrm{(erg/s)/(M_{\sun}/yr)} $ \citep{hmxb}.

\section{Conclusion}

We investigated the origin of unresolved X-ray emission
from M31 using Chandra and XMM-Newton data.  
We demonstrated that it consists of three different components:

\begin{enumerate}

\item
Broad-band emission associated with old population, similar to the
Galactic ridge  emission in the Milky Way. 
It is a a combined emission of a large number of weak unresolved
sources of stellar type, the main contribution being from cataclysmic
variables and active binaries. The surface brightness distribution of
this component approximately follows the distribution of K-band
light. The absorption corrected X-ray to K-band luminosity ratios are
compatible with the Milky Way values. The total luminosity of 
this component inside central $800 \arcsec\times400 \arcsec$ is of the order of
$\sim 3 \times 10^{38}$ erg/s in the $0.5-10$ keV band. 

\item
Soft emission localized in the inner bulge of the galaxy along its
minor axis. This
emission is from ionized gas with the  temperature of the order  
of $\sim 300$ eV, although its spectrum can not be adequately described
by a simple one- or two-temperature model of optically-thin emission
from a gas in a collisional ionization equilibrium. The $0.5-2$ keV 
luminosity  in the central $8 \arcmin \times 20\arcmin$ area is
$\sim 2\times 10^{38}$  
erg/s, the absorption corrected bolometric luminosity is  
$\sim 5\times 10^{38}$ erg/s. The total mass of the gas is 
$\sim 2\times 10^{6} \ \mathrm{M_{\sun}}$, its cooling time $\sim 250$
Myrs. 
The surface brightness distribution is drastically different from the
stellar light distribution, it is significantly elongated along
the minor axis of the galaxy. The morphology of the soft emission
suggests that gas outflows along the direction
perpendicular to the galactic plane. The mass and energy
budget is maintained by the mass loss by the evolved stars and type Ia
supernovae. The ``vertical'' extent of the gas exceeds $\ga 2.5$ kpc.
These results are in good agreement
with those recently obtained by \citet{li}.

\item
Hard emission from spiral arms. In the 0.5--7 keV band this emission
has approximate power law shape with $\Gamma\approx 2$. It is most
likely associated with star-forming regions and is due to 
young stellar objects and young stars.
The $ \mathrm{L_{X}/SFR} \approx 9.4 \times 10^{38} \
\mathrm{(erg/s)/(M_{\sun}/yr)} $ which is about $\sim 1/3$ of the
contribution of HMXBs.
\end{enumerate}

\bigskip
\small
\noindent
\textit{Acknowledgements.}
We thank the anonymous referee for useful and constructive comments.
This research has made use of Chandra archival data provided by the Chandra 
X-ray Center. XMM-Newton is an ESA science mission with instruments 
and contributions directly funded by ESA Member States and the USA (NASA).
This publication makes use of data products from Two Micron All 
Sky Survey, which is a joint project of the University of Massachusetts and 
the Infrared Processing and Analysis Center/California Institute of Technology,
funded by the National Aeronautics and Space Administration and the 
National Science Foundation. The Spitzer Space telescope is operated by
the Jet Propulsion Laboratory, California Institute of Technology, under 
contract with the National Aeronautics and Space Administration. 

{}

\end{document}